# The Linear Momentum as a Tensor Function.

Enrique Ordaz Romay[1]

*Facultad de Ciencias Físicas, Universidad Complutense de Madrid*

# Abstract

At present, whenever we work in newtonian mechanics we consider momentum to be a three-dimensional vector or a 4-dimensional one when we work in relativistic mechanics. However, this mathematical vector model has barely 200 years and its complete installation in the physics is hardly more than 100 years old.

In classical mechanics, based on point particles with mass, momentum is considered as a vector because the speed of the particle is applied on a single point and the mass is defined only at that point. When the extension to classical rigid solids is made, that is, to solids that keep their shape, being so that the speed is defined to be the same for each and every point where the solid is defined, then the viewpoint of momentum as a vector is the simplest and more convenient approach.

Relativity (restricted and general) was developed in the decades of 1900 and 1910 and quantum mechanics in the decades of 1920 and 1930. Relativity then impedes the existence of classical rigid solids. The maximum speed for the transmission of signals prevents a body from varying its status without being deformed. On the other hand, quantum physics breaks up with the concept of point particles with mass, substituting this model for the one of the wave function defined in an extensive region of the space (space-time in the quantum relativity). Can we really define, the moment of a relativistic solid or of a wave function with a simple vector?

---

[1] enorgazro@cofis.es

# Introduction [1]

Momentum is a physical - mathematical concept whose origin goes back to the times of the ancient Greece. Aristotle (384–322 A. D.) spoke about the movement of a body as an inherent quality of the own object in function of its nature. In this way, the nature of a stone is in the soil and for this reason the stone falls. Aristotle distinguishes among the quantity of movement that two bodies can develop in function of their different masses.

This primitive concept of quantity of movement will be approached to by some authors of the Middle Ages who, in turn, will introduce a new concept: impulse or impetus. The impulse became the quality that the propelling instrument gives to the object so that it maintains its movement. Then, it was regarded as a secondary cause.

Galileo Galilei (1564–1642) will be the first one to study the concept of inertia experimentally, using ramps or inclined planes. Galileo observed that a moving body which has descended along an inclined plane, ascends along another one back to the initial height regardless of the inclination of the second ramp. Thus, he concludes that if the second plane is horizontal and there would be not friction, the body would continue moving indefinitely. This is one form of the principle of inertia.

Finally, in the XVII century René Cartesius (1596-1650) unifies the dispersed concepts of quantity of movement, momentum, inertia, impulse, etc. and includes them in only one concept to which he simply calls momentum and to which he identifies with the expression $p = m \times v$ (momentum is equal to the mass of an object times its speed). Unfortunately Cartesius got lost in metaphysical reasoning and he did not study in depth the physics point of view.

Later, Isaac Newton (1642 – 1727), in his second law of the mechanics established the form of the force like $F = m \times a$, but it will not be until Leonhard Euler

(1701-1783) that the differential relationship $F = \frac{dp}{dt}$ can be established. On the other hand, Euler becomes aware of the importance of the kinetic energy *T* and he established its relationship with the momentum as: $p = \frac{\partial T}{\partial v}$.

From the substitution of these two expressions, and knowing that when a force comes from a potential *U*, the form of this force is: $F = -\frac{dU}{dx}$ (for the one-dimensional case), the second law of Newton takes the form: $-\frac{dU}{dx} = \frac{d}{dt}\left(\frac{\partial T}{\partial v}\right)$. This expression is the starting point of the analytical physics of Joseph-Louis Lagrange (1736–1813). For a coordinates change such as *q=q(x)* the previous expression takes the form: $-\frac{dU}{dq} = \frac{d}{dt}\left(\frac{\partial T}{\partial \dot{q}}\right)$. This simplifies the mathematical operations, because we can work directly with generalized coordinates. In such coordinates the form of momentum will be: $p = \frac{\partial L}{\partial \dot{q}}$, being: *L = T - U* the function of Lagrange.

Then, they will be Daniel Bernoulli (1700–1782) and Jean le Rond d'Alambert 1717–1783) the ones who will introduce the vector calculus in physics, although vector calculus will be introduced in the different branches of physics by different authors and at different times until the arrival of the beginning of the XX century. The reason to use vector calculus is to make the expressions more compact and intuitive. In its components form, the momentum will have the form $p_\alpha = m \cdot v_\alpha$, while in vector notation it will be: $\vec{p} = m \cdot \vec{v}$.

Albert Einstein's relativity (1879–1955) passes from the three-dimensional vector calculus to four-dimensional and while increasing the number of indexes he passed the expressions from vectorial to tensorial. The three-dimensional momentum vector becomes the energy-momentum four-vector. The reason is the following one: energy, in analytic mechanics and in Cartesian coordinates has the form $E = \vec{p} \cdot \vec{v} - L$, being so that $p = \gamma \cdot m \cdot \vec{v}$ and $L = -\gamma^{-1} \cdot m \cdot c^2$; ($\gamma = 1/\sqrt{1 - v^2/c^2}$) substituting we are left with $E = \gamma \cdot m \cdot c^2$. It is easy to see that $E^2 - p^2 c^2 = m^2 c^4$. This induces us to think that energy (divided by *c*) is the "time" component of a four-vector whose space components are those of the linear momentum. This is endorsed by the analytical mechanics thanks to the fact that, when the action *S* is a function of the coordinates $p_i = -\frac{\partial S}{\partial q_i}$ [2], while $H = -\frac{\partial S}{\partial t}$ (*H* is the function of Hamilton that, in mechanics, and for non-rotational forces, represents the energy of the system), allowing therefore that $p_0 = -\frac{E}{c}$.

Quantum mechanics gives a new expression to the momentum. Following the principle of complementarity the components of the four-vector energy-momentum have the form $p_i(\Psi) = -i\hbar \frac{\partial \Psi}{\partial x_i}$. This substitution maintains the relativistic invariance, being in this way compatible with the relativity.

Wolfgang Pauli (1900 - 1958) introduces the concept of spin in the mathematical formalism as the product of the traditional scalar wave function $\Phi$ by a vector $s_\alpha$ (called spinor), consequently getting the expression of the total wave function to have the form: $\Psi_\alpha = \Phi \cdot s_\alpha$. Soon afterwards, the quantum relativism of Paul Dirac (1902–1984) demonstrates that the vector or tensor construction of the wave function is something innate with the actual relativist treatment of quantum mechanics. It only changes where, instead of 2*n*+1 components of the spin vector for each state of spin *n*, what we get is 2*n* indexes of 4 components each one, for each state of spin *n*.

The last step given by Pauli and Dirac implies, by substitution in the principle of complementarity, that the momentum can be a tensor [3]. Why did the physics never give this last step? What sort of physical meaning may have something like a tensor momentum?, How should other mathematical expressions change their range?.

In this article we try to find some answers to this questions.

# Magnitudes and physical qualities

The physical concepts used in physical theories are of two types [4]:

- Physical magnitudes: they have material reality, as the mass or the speed.

- Physical properties: they are qualities, behaviours, descriptions an so on but, in principle, they do not have material physical reality they are things like inertia, gravity (as a concept) and force fields[2].

It is quite easy to confuse those abstract properties with the realities of the physical magnitudes they qualify to, and to which, due to the constant use of the language, they seem to represent. As an example, let us speak about the property we call "inertia".

"Inertia" is a quality of matter according to which every material object, in absence of forces, tends to remain in its state of rest or movement. The concept of inertia is the description of a particular behaviour of the object, but it is not itself a physical entity. The "inertia" does not have, then, its own material entity. It can be explained, demonstrated or be used as cause or consequence of other physical properties, but this does not mean it can become something material. It is only the description of an observed behaviour.

In this sense, the mathematical form by which we implement the inertia will be adjusted to the needs of the mathematical formalism that we use, as well as to the form of the different physical laws.

---

[2] We say that in principle they do not have material physical reality because they are products or constructions elaborated by the mind and, merely this does not make them necessarily real. However, some of these may have physical reality, after all.

# Quantum physics and the observables.

In quantum physics only observables applied on wave functions have material entity, but not the observables themselves. When we say that, according to the principle of complementarity, the momentum has the form: $p_i = -i\hbar \frac{\partial}{\partial x_i}$ [5], we are not providing any material fact or the result of any observation. Now then, when one says that the momentum of a physical system represented by the wave function $\Psi$ is $p_i(\Psi) = -i\hbar \frac{\partial \Psi}{\partial x_i}$, we are certainly giving the result of a direct physical observation, that is, a value for a magnitude. In this case, the result is a vector function. If a numeric (vector) result were desired for the value of the momentum, it would be necessary to calculate the auto-values of $p_i(\Psi)$. If we want to know the probabilities of each value, then it would be necessary for us to decompose $\Psi$ in the auto-functions of $p_i(\Psi)$ and to calculate the amplitudes….

# The field theories

The field theories (such as relativistic quantum mechanics) substituted the concepts of particle, object and matter by force fields, potential energy and space-time metrics [6]. Analytic mechanics is applied both to the theories of particles and to the fields ones and due to this, the Lagrangian, Hamiltonian, generalized momentum, etc, can refer both to material objects and to force fields.

However, a great difference exists between, for example, the momentum of a particle (point) and of a field (like the electromagnetic): while the material particle has a traditional momentum as a function of its mass and its speed, the field does not contain in itself neither mass or speed. When we speak about the momentum of a field we suppose that we are applying the field to a particle, that is to say, our variables are the parameters that define the point-particle. However, this does not guarantee that the momentum of the field (before being applied to the particle) is necessarily a vector.

The mass of a particle is a measurable physical magnitude, with a numeric value (scalar). In this way, the speed of a particle is another physical magnitude that does not only have a numeric value, but also an application point (the position of the particle) and a direction. That is to say: it is a traditional vector. This is why, together with the rules of the vector calculus (according to which the product of a scalar by a vector is another vector whose components are the first ones multiplied by the scalar), the momentum of a particle, defined by the product of the mass of the particle by its speed, is also a vector.

This fact does not change from the point of view of the analytical mechanics for point particles with mass. Since $p_i = -\partial S / \partial x_i$, for each value of the index $i$, we have one component of the 4-momentum.

However, when we are in theory of fields, the only thing that we know, a priori, is that, once the particle (point) is applied with its parameters, in the field, the result should be a vector. This condition can be fulfilled under many different forms. Two examples are:

- In absence of field, the 4-vector speed has the form $u^i = \dfrac{1}{c\sqrt{1-\dfrac{v^2}{c^2}}} \begin{pmatrix} c \\ v_x \\ v_y \\ v_z \end{pmatrix}$ and the momentum is calculated by: $p^i = mc \cdot u^i$ [6]. In this case, the momentum of the system is represented by a scalar: the mass of the particle times the light speed.

- In a homogeneous but not isotropic space, the momentum of a particle can have the lineal form: $p^i = M(m)^i_j \cdot u^j$. In this case, the momentum of the field has a matrix form.

The cause of the difference is that, momentum, just as we understand it, is deeply connected to the concept of point particle with mass, and this it is one cornerstone of the whole classical physics. If we wonder what is the momentum of a rigid solid we usually consider the momentum of their mass centre as a valid answer. But this is not exact, the momentum is distributed in the whole surface of the solid and it depends of the angle that forms each facet of the surface with the vector speed. That is to say, for a rigid solid, the momentum is a distribution function that depends of more factors, besides the mass and speed of the object. The momentum of the rigid solid is not then, a vector.

From the point of view of the fields theory, once we apply a particle to the field, with its position, mass, charge, speed, etc., we can specify the value of all the traditional magnitudes, as the momentum, and therefore we can return to the traditional mechanics of point-like matter.

However, from the viewpoint of a pure field theory, such as the frame of the relativistic quantum mechanics, which is the mathematical form of physical concepts just as momentum, action, the Lagrangian or the Hamiltonian?

To summarize: the momentum of an individual particle (a point with mass) is a vector. However, the momentum of a force field or a particle that is not a point (because it can be defined by a distribution function) can take the form of any mathematical expression, as long as whenever it is applied to a specific particle (a point with mass), or, after the necessary approximations are made to consider the particle as a point (assuming, for example that the interval in which the distribution function has a non-zero value, it is insignificant in relation with the system) the result is a vector. In this context, which is the mathematical form of the physical concepts just as the momentum, the action, the Lagrangian or the Hamiltonian?.

## Tensor magnitudes

To respond to the previous question, the easiest concept to be tackled, from the quantum point of view is the momentum because the principle of complementarity gives us the following answer: as $p_i(\Psi) = -i\hbar \dfrac{\partial \Psi}{\partial x_i}$ [5], then when $\Psi$ is scalar $p_i(\Psi)$ will be vector, but when $\Psi$ has *n* indexes, $p_i(\Psi)$ will have *n* + 1 indexes.

But this easiness does not indicate that this fact is exclusive of the quantum theory. In classical theory of fields we have the case of a free particle. Its action can be expressed in the form [6]:

$$S = \int_b^a -mc \cdot ds$$

Knowing that $ds^2 = dx_i dx^i$ we can then observe that the action *S* is the norm of an action tensor (vector) of the form: $S_i = \int -mc \cdot dx_i$ .

# Meaning of the tensor action and tensor momentum in quantum mechanics

What is the meaning of a tensor action or a tensor momentum? This question could have a simple answer from the point of view of the quantum mechanics. If the wave function is tensorial, this is due to a quantum effect called spin. As in the traditional physics a quantum spin is never observed what we should do though, is to work, in those cases in which the spin is not of interest or it is insignificant, with the norm of the wave function.

In fact, we usually do that unconsciously. We know that, for particles with spin ½, the wave function is vectorial (spinorial), so that, we divide the wave function in two components: a scalar and a vector one with its norm equal to the unit: $\Psi_\alpha = \Phi \cdot s_\alpha$ with $\Phi = |\Psi|$ and therefore $|s_\alpha| = 1$. We work with the scalar part scalar while we do not need the spin, and whenever we need it, we work with the spin component directly.

However, this reasoning involves a serious defect: the spin is not simply a quantum effect of which a statistical balance can be made so that its effects are not noticeable in macrosystems. The spin is the result of applying the quantum physics to the relativistic structures.

For the quantum physics a particle is a duality of wave and corpuscle, that it is distributed in the whole space-time. However, in relativity applied to matter just as we understand it in macrosystems, only the point particles have meaning [6]. In fact, that is the origin of the paradoxes produced by the famous EPR theorem [7].

Just as we saw earlier, a momentum vector cannot describe the momentum of a rigid solid, because it is not a point particle, in relativistic quantum mechanics, the momentum tensor comes to tell us this same thing: the particles described by wave functions are not point-like. In fact, whenever we apply the concepts of linear or vector momentum, scalar action, Lagrangian, or Hamiltonian or any other physical quantity that is derived from the concept of point particle, we will find problems of undefinitions that will only be sorted out by mathematics "recipes" such as the Wick product [8].

The physical meaning of a linear momentum tensor, in quantum physics, is deduced directly from the physical meaning of a wave function that is not scalar. The components of spin of a wave function are due to fact that the solution of the relativistic wave equation does not have representation in a Hilbert scalar space. Then it needs the Cartesian product of $4^{2n}$ (being $n$ the spin) spaces of Hilbert in order to be represented. To say it in a more intuitive form: a particle of spin ½ cannot be represented by a point in a Hilbert space, but, certainly, by a 4-vector in a 4-space of Hilbert. Just as the momentum of a particle of spin 1, having two indexes, cannot be represented as a 4-vector in a 4 space of Hilbert, but as a matrix in a 4×4 space of Hilbert.

Another way of explaining it: a particle of ½ spin, being represented by a vector with four components, comes to be represented by "the Cartesian product of 4 particles of zero spin (scalars), each one in a Cartesian axis (of Hilbert)." As for each of the particles of zero spin that represents to our particle of ½ spin, we can calculate a momentum vector with four components, the particle of ½ spin can define 4 linear momentum vectors with 4 components each, or what it is the same: a momentum matrix with 4×4 components. As long as the space is a plane (a space of Minkowski) the four momentum vectors of each particle will be parallel, but what does it happen when the space is not flat, as in the case of the gravitational field?

## Momentum in classical mechanics

Let us pass to the case of the classical mechanics. In this frame, the momentum of an extensive body is usually expressed by the product of the speed of its mass centre by the mass of the solid. However, this is terribly inexact unless we make some approximations such as assuming that the dimensions of the body are negligible and/or that the body is a rigid solid and/or the crashes are elastic or etc, etc.

The traditional form of the momentum of a body corresponds to the "sum" of each one of the momentums of the components that form the body. When the body is usually approximated or simplified to a rigid solid it is being said that the speeds, of each one of those components, are equal in magnitude and direction. Therefore, the sum is reduced to the sum of the masses:

$$\vec{p} = \sum m_i \vec{v}_i \text{, as } \vec{v}_i = \vec{v} \text{ for all } i \text{, it is } \vec{p} = \left(\sum m_i\right)\vec{v} = M \cdot \vec{v}$$

This classical calculation is based on the assumption that the speeds of each of the component parts of the body are all the same. However, in general, the speeds of the different parts of a solid depend of the position that such a part occupies inside the body. The relativity tells us that a minimum time delay exists for an alteration to be transmitted from one point of the body to another, therefore, the speed of each point of a body, even in the case of a rigid solid, cannot be the same if the body does not maintain an uniform speed.

Let us suppose that a body is defined by the distribution function of its mass in the three space coordinates, that is to say, by its function of mass density ρ(x, y, z). The speed of each component is defined by means of a vector function for each point of the space in which the body is defined. It assigns a vector value representing the speed in that point $\vec{v} = \vec{v}(x, y, z)$. It will always be possible to decompose this function as the sum:

$$v_\alpha = v_\alpha(0) + \frac{\partial v_\alpha}{\partial x_\beta} x_\beta + ... = v_\alpha(0) + (\partial_\beta v_\alpha) x_\beta + ...$$

(We use the convention of Einstein for the sum and for the repeated indexes, although at the moment, the calculations are not relativistic). The meaning of $v(0)$ is the speed of the mass centre. Since we are more interested in the variations of the speed than in its constant part, we could always choose a system of reference, bound to the body, in which this speed is equal to zero. However, we will not do it because, in this way, we can obtain, at the end, a more intuitive approximation to the classical case.

On the other hand, the mass of the body will decompose in the sum of its parts. Each part will have a mass (this is valid for both continuous as discontinuous bodies, it only depends on the form of the density):

$$dm = \rho(x_1, x_2, x_3) \cdot dx_1 \cdot dx_2 \cdot dx_3$$

Consequently, The momentum of each one of the parts of the body will have the form:

$$p_\alpha = (\rho \cdot v_\alpha(0) + \rho \cdot (\partial_\beta v_\alpha) x_\beta) \cdot dx_1 \cdot dx_2 \cdot dx_3 + ...$$

The momentum of the whole body is made with the sum of each and all of its parts. Therefore, integrating, and keeping in mind that $v(0)$ is constant, we obtain:

$$p_\alpha = m \cdot v_\alpha(0) + \iiint_\infty \rho(x_1, x_2, x_3) \cdot (\partial_\beta v_\alpha) x_\beta \cdot dx_1 \cdot dx_2 \cdot dx_3 + ...$$

When the body is a classical rigid solid, the partials of the speeds are zero and the second term is nulled (or it becomes constant). This only leaves us with the first term and it identifies, as we said earlier, to $v_\alpha(0)$ with the speed of the mass centre. Nevertheless, this integral will not be equal to zero elsewhere than in systems with uniform speed. For any other ones, we will have to calculate the value of this integral.

Thus, in general, the momentum of a body is a function of the coordinates of the part of the body where we measure it (just like the speed of the body). In the reference system centred in the mass centre we can see that (we go beyond from the Greek sub-indexes to the Latin ones to highlight that, from now on, the calculations are in the four space-time coordinates), in the first approximation we have:

$$v_i = (\partial_k v_i) x^k \quad p_i = (\partial_k p_i) x^k$$

For a body with extension, considered as a whole, the only parts of these two expressions that make any sense are those among parenthesis. That is to say: the Jacobiants. The Jacobiant of the speed and of the momentum are the only two expressions that do not make reference to a concrete point of the space, but to the set of points where the body is defined.

On the other hand, given the linearity of these expressions, any property of the vector function is also inherited by the Jacobiant. Therefore, the function:

$$\boxed{P_{ik} = \partial_k p_i}$$

Behaves as the momentum of an extensive body.

Now, the units of the momentum will depend of the number of indexes that this momentum contains. This way, the momentum vector that represents a point particle with mass will have the units of Mass×Length×Time$^{-1}$. On the other hand, when the number of indexes is two, the units will be those of Mass×Time$^{-1}$. Similarly, with three indexes these units will be then: Mass×Length$^{-1}$×Time$^{-1}$ and so on.

These units have certain asymmetry. However, this asymmetry is caused by the use of units that do not take into account the relativistic considerations, as well as by the use of an ordinary speed vector instead of the 4-dimensional speed vector.

In relativity, the following form defines the 4-dimensional speed vector [6]

$$u^i = \frac{dx^i}{ds} = \gamma \cdot \left(1, \frac{\vec{v}}{c}\right)$$

It is easy to observe that its units are null.

Now then, (continuing in relativity) in a no point-like body (extensive body) in which the speed is not a vector in itself, but rather it will be so, depending of its position in the space, (that is to say, for $V_{ij}$, being $V_i = V_{ij} x_j$), the speed of the body $V_{ij}$ will have units of Length $^{-1}$. In general, a system in which the speed is defined by $n$ indexes, the units of such speed will be of Length $^{-n+1}$.

In this way, the units of the momentum, when it has $n$ indexes, will be: Mass× Speed×Length$^{-n+1}$, matching with the traditional relativistic definition of the momentum 4-vector whose form is: $p^i = mc \cdot u^i$ and consequently its units are those of Mass×Speed.

## Conclusion

Momentum is a physical–mathematical concept which definition depends of the branch of physics in which we are focusing. Traditionally, this physical model was the one of "point-particle with mass" or the one of "rigid solid", both concepts deeply anchored in the newtonian physics. In both of these cases, the physical model represents the speed by a vector and the mass by a scalar. The result is that the momentum has always been represented as a vector.

Still, this reduction or simplistic vision of the newtonian physical systems has been carried to all of the other branches of physics, both in field theories and in relativity or quantum physics, where the concept of a point-particle with mass, does not make sense or generates paradoxes.

"A quantum–relativistic physical system will never be well defined if we use the same classical and intuitive concepts of the point-particles with mass or of the rigid solids." This is an idea all of us know well, but do we really apply it? And, if we do not dare to properly define the quantum–relativistic physical systems, how can we find a unique theory to describe these systems?

## Acknowledgement:

Thanks to Jorge Millán for takes an interest, for his help and for revise the last version of this paper.

## Bibliography.